\documentclass[aps,prd,reprint,superscriptaddress,nofootinbib]{revtex4-2}
 \usepackage{amsmath}
 \usepackage{amssymb}
 \usepackage{subfigure}
\usepackage{tensor}     
\usepackage{graphicx}   
\usepackage[
colorlinks=true,        
citecolor=blue,         
linkcolor=blue,         
urlcolor=blue           
]{hyperref}             
\usepackage{bm}         
\usepackage{xcolor}     
\usepackage{tabulary}
\usepackage{color}      
\usepackage[utf8]{inputenc} 
\usepackage[section]{placeins} 
\usepackage{graphics,appendix,afterpage,makecell} 
\usepackage{orcidlink}
\begin{document}
	
\title{Non-minimal coupling in light of ACT} 

\author{Qing Gao}
\email{gaoqing1024@swu.edu.cn}
\affiliation{School of Physical Science and Technology, Southwest University, Chongqing 400715, China}

\author{Yungui Gong \orcidlink{0000-0001-5065-2259}}
\email{gongyungui@nbu.edu.cn}
\affiliation{Institute of Fundamental Physics and Quantum Technology, Department of Physics, School of Physical Science and Technology, Ningbo University, Ningbo, Zhejiang 315211, China}

\author{Zhu Yi \orcidlink{0000-0001-7770-9542}}
\email{Corresponding author. yz@bnu.edu.cn}
\affiliation{Faculty of Arts and Sciences, Beijing Normal University, Zhuhai 519087, China}
\affiliation{Advanced Institute of Natural Sciences, Beijing Normal University, Zhuhai 519087, China}

\author{Fengge Zhang}
\email{zhangfengge@hnas.ac.cn}
\affiliation{Institute for Gravitational Wave Astronomy, Henan Academy of Sciences, Zhengzhou, Henan 450046, China}


\begin{abstract}
The latest ACT data release disfavors the attractor $n_s=1-2/N$. 
In inflationary models with nonminimal coupling, such attractors typically arise in the strong coupling limit. 
To align with observational constraints, we focus on nonminimal coupling models with small coupling constants.
For the model with the coupling function $\Omega(\phi) = 1 + \xi f(\phi)$ and the potential $V(\phi) = \lambda^2 f^2(\phi)$, 
we find that observational data constrain the parameters as $0.1 \lesssim \xi \lesssim 35$ and $0 \lesssim k \lesssim 1.5$ for $f(\phi) = \phi^k$ at the  $1\sigma$ confidence level. 
With the help of the nonmiminal coupling $\Omega(\phi) = 1 + \xi \phi^2$, 
the hilltop inflation and power-law inflation models with power indices $2/3$ and $1/3$ can be consistent with observational data within the  $1\sigma$ range. 
We also give the viable parameter regions for $\xi$ for these three models.
\end{abstract}
	
\maketitle
\section{introduction}
Inflation elegantly resolves several fundamental problems in the standard big bang cosmology, including the flatness, horizon, and monopole problems, while simultaneously providing the initial conditions for the formation of large-scale structure in the universe~\cite{Starobinsky:1980te, Guth:1980zm, Linde:1981mu, Albrecht:1982wi}. Quantum fluctuations of the inflaton field generate metric perturbations, which leave characteristic imprints as temperature anisotropies in the cosmic microwave background (CMB).

The inflationary attractor $n_s = 1 - 2/N$ for a broad class of inflationary models, 
such as hilltop inflation \cite{Boubekeur:2005zm}, T model \cite{Kallosh:2013hoa}, E model \cite{Kallosh:2013maa},  $R^2$ inflation \cite{Starobinsky:1980te} and Higgs inflation with nonminimal coupling $\xi\phi^2 R$ in the strong coupling regime ($\xi \gg 1$) \cite{Kaiser:1994vs, Bezrukov:2007ep},
lies near the central value $0.9649$ of the {\it Planck} 2018 measurements on CMB \cite{Planck:2018jri} for $N=60$, 
where $N$ denotes the number of $e$-folds before the end of inflation. 

Besides $R^2$ inflation and Higgs inflation with nonminimal coupling, the so-called universal attractor $n_s = 1 - 2/N$ can also be obtained in a more general nonminimal coupling model. 
In this model, the coupling takes the form $\xi f(\phi) R$ and the inflation potential is $\lambda^2 f^2(\phi)$ for arbitrary $f(\phi)$ in the strong coupling limit~\cite{Kallosh:2013tua}. 
Choosing $f(\phi) = \phi^2$ recovers the Higgs inflation scenario. For these non-minimal coupling inflation models, the universal attractor  $n_s = 1 - 2/N$  arises under the strong coupling condition. 

Recently, new data from the Atacama Cosmology Telescope (ACT) have been released \cite{Louis:2025tst,ACT:2025tim}, 
indicating a higher value of $n_s$ compared to previous {\it Planck} constraints. 
A joint analysis of {\it Planck} and ACT data (P-ACT) yields $n_s = 0.9709 \pm 0.0038$ \cite{Louis:2025tst}. 
By incorporating additional data from CMB lensing and the Baryon Acoustic Oscillation (BAO) data measured by the Dark Energy Spectroscopic Instrument (DESI)  \cite{DESI:2024uvr, DESI:2024mwx} (P-ACT-LB), 
the value further increases to $n_s = 0.9743 \pm 0.0034$ \cite{Louis:2025tst}, 
deviating from the original Planck result by about $2\sigma$. 
Notably, the attractor $n_s = 1 - 2/N$ 
is disfavored by this latest P-ACT-LB constraint, and located at the regions of constraints $2\sigma$~\cite{ACT:2025tim}.

To reconcile the wide classes of inflationary attractors with current observations, 
it is necessary to consider nonminimal coupling inflationary models where the attractors are not reached or
the strong coupling limit is relaxed.  
For instance, with a  value of  $\xi = 1$ and $f(\phi) = \phi$, the predicted value becomes $n_s = 1 - 2/(3N)$, which aligns with the new P-ACT-LB constraints for $N=60$ \cite{Kallosh:2025rni}.
For additional approaches to address the implications of the P-ACT-LB results, see Ref. \cite{Gialamas:2025kef,Frob:2025sfq,Dioguardi:2025vci,Brahma:2025dio,Berera:2025vsu,Aoki:2025wld,Dioguardi:2025mpp,Salvio:2025izr}.

Since a nonminimal coupling helps to reduce the value of the tensor-to-scalar ratio $r$, in this paper, we investigate inflationary models with nonminimal coupling where the coupling constant is small, 
focusing on their ability to account for the latest P-ACT-LB constraints on the scalar spectral index.  
The paper is organized as follows.
In Sec. II, we discuss nonminimal inflationary models under the strong limit condition, i.e., the universal  attractor. 
The nonminimal inflationary models in the general case is discussed in Sec. III.  The conclusions are drawn in Sec. IV. 
We set the unit by  $M_{pl}=(8\pi G)^{-1}=c=1$, where $M_{pl}$ is the reduced Planck mass and $c$ is the speed of light.

\section{Nonminimal Inflation Model: Strong Coupling Limit}

The action for a general scalar-tensor inflation model in Jordan frame is given by
\begin{equation}
\label{jscten}
S = \int d^4x \sqrt{-\tilde{g}} \left[\frac{1}{2}\Omega(\phi)\tilde{R}(\tilde{g}) - \frac{1}{2}\omega(\phi)\tilde{g}^{\mu\nu} \nabla_{\mu}\phi\nabla_{\nu}\phi - V_J(\phi)\right],
\end{equation}
where $\Omega(\phi)$ and $\omega(\phi)$ are coupling functions, and $V_J(\phi)$ is the potential. In this work, we restrict to a canonical kinetic term, i.e., $\omega(\phi) = 1$.  Applying the following conformal transformations,
\begin{gather}
\label{conftransf1}
g_{\mu\nu} = \Omega(\phi) \tilde{g}_{\mu\nu}, \\
\label{conftransf2}
d\psi^2 = \left[\frac{3}{2} \frac{(d\Omega/d\phi)^2}{\Omega^2(\phi)} + \frac{\omega(\phi)}{\Omega(\phi)} \right] d\phi^2,
\end{gather}
the action \eqref{jscten} is brought into Einstein frame:
\begin{equation}
\label{escten}
S = \int d^4x \sqrt{-g} \left[ \frac{1}{2}R(g) - \frac{1}{2}g^{\mu\nu} \nabla_\mu\psi \nabla_\nu\psi - U(\psi) \right],
\end{equation}
where the potential in Einstein frame is 
\begin{equation}\label{potential:rel}
  U(\psi) = \frac{V_J(\phi)}{\Omega^2(\phi)}. 
\end{equation}
In the strong coupling limit,
\begin{equation}
\label{strcouplim1}
\frac{3}{2}\frac{[d\Omega(\phi)/d\phi]^2}{\Omega^2(\phi)}\gg \frac{\omega(\phi)}{\Omega(\phi)},
\end{equation}
we get
\begin{equation}
\label{psiphi1}  
\psi\approx \sqrt{\frac{3}{2}}\,\ln\Omega(\phi),
\end{equation}
and
\begin{equation}
\label{psiphi2}  
U(\psi)\approx V_J(\phi)\exp\left(-\sqrt{8/3}\,\psi\right).
\end{equation}

For the Higgs potential,
\begin{equation}\label{Higgs:potential}
    V_J(\phi) = \frac{\lambda}{4} \phi^4,
\end{equation}
with minimal coupling, $\Omega(\phi) = 1$, 
the predicted tensor-to-scalar ratio $r$ is too large to be consistent with observational data. To obtain a smaller $r$, 
nonminimal coupling is considered. With $\Omega(\phi) = 1 + \xi\phi^2$ and the Higgs potential \eqref{Higgs:potential}, 
the so-called Higgs inflation model~\cite{Bezrukov:2007ep} is described by
\begin{equation}
\label{higgs:action}
S = \int d^4x \sqrt{-\tilde{g}} \left[ \frac{1}{2}(1+\xi\phi^2) \tilde{R}(\tilde{g}) - \frac{1}{2}\tilde{g}^{\mu\nu}\nabla_{\mu}\phi\nabla_{\nu}\phi - \frac{\lambda}{4}\phi^4 \right].
\end{equation}
In the strong coupling limit, 
\begin{equation}
    \phi \gg 1/\sqrt{\xi},
\end{equation}
and $\xi\gg 1/6$,
the potential in Einstein frame becomes
\begin{equation}
\label{hissen1}
 U(\psi) = \frac{\lambda}{4\xi^2} \left[1+\exp(-2\psi/\sqrt{6})\right]^{-2},
\end{equation}
and the predictions for the scalar spectral index and tensor-to-scalar ratio are
\begin{equation}
\label{attractor:nsr}
n_s - 1 = -\frac{2}{N}, \quad r = \frac{12}{N^2}.
\end{equation}
The potential \eqref{hissen1} coincides with that of the Starobinsky model~\cite{Starobinsky:1980te} in Einstein frame. For $N=60$, the predictions become
\begin{equation}\label{prediction:higgs}
    n_s = 0.967, \quad r = 0.003,
\end{equation}
which are compatible with the $1\sigma$ confidence region from {\it Planck} 2018 data,
but are disfavored by the $1\sigma$ regions of the recent P-ACT-LB results. 

The Higgs inflation model \eqref{higgs:action} is a special case of the universal attractor model~\cite{Kallosh:2013tua}, in which the coupling function and potential are
\begin{equation}
\label{attractor:act}
\Omega(\phi) = 1 + \xi f(\phi), \quad V_J(\phi) = \lambda^2 f^2(\phi).
\end{equation}
By choosing $f(\phi) = \phi^2$, Higgs inflation is recovered. Under the strong coupling limit,
\begin{equation}
    \Omega(\phi) \ll \frac{3}{2}\Omega'(\phi)^2,
\end{equation}
for general $f(\phi)$, the Einstein frame potential reduces to the universal attractor form, yielding the predictions of Eq.~\eqref{attractor:nsr}. As with the Higgs and Starobinsky models, these predictions are now disfavored by the latest observational data.

Thus, to remain consistent with current observations, the strong coupling limit must be relaxed.

\section{Nonminimal coupling models}

When the strong coupling condition is not imposed, the predictions for $n_s$ and $r$ in the universal attractor model~\eqref{attractor:act} depend on both the coupling parameter $\xi$ and the form of $f(\phi)$. 
It is therefore possible to achieve consistency with the P-ACT-LB data for small coupling value of $\xi$. 

\subsection{General coupling function}

\begin{figure*}[htpb]
    \centering
    \includegraphics[width=0.49\linewidth]{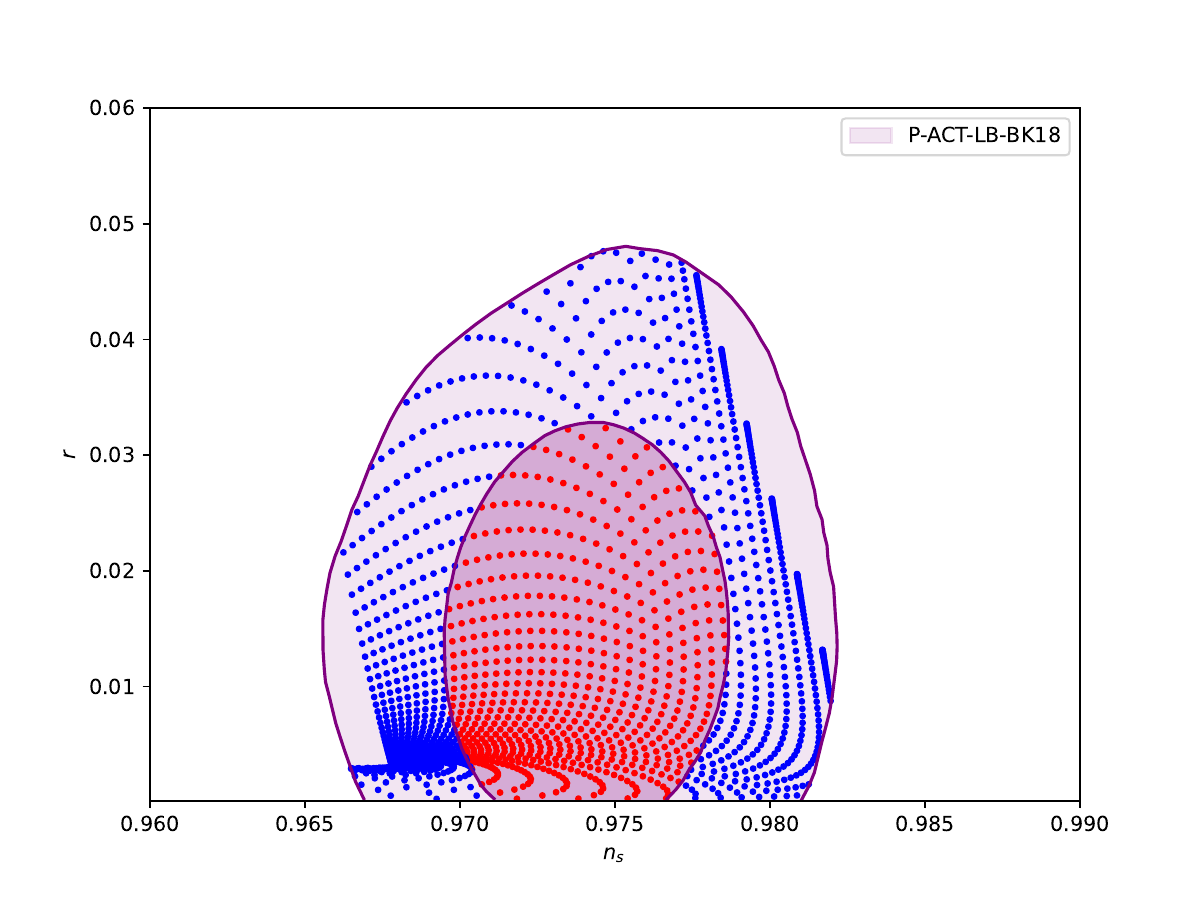}
    \includegraphics[width=0.49\linewidth]{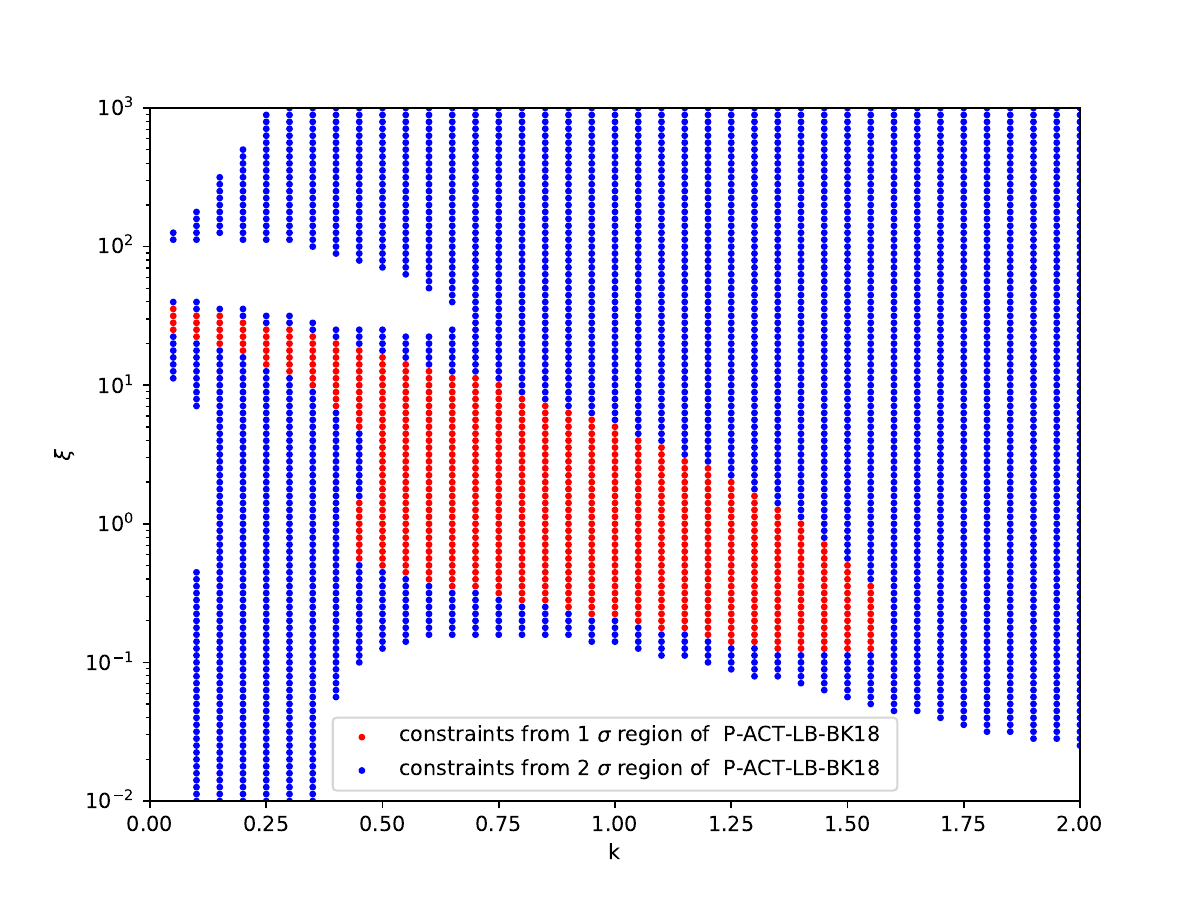}
    \caption{(Left) Scalar spectral index $n_s$ and tensor-to-scalar ratio $r$ predicted by the nonminimal coupling inflation model for various values of $k$ and $\xi$, compared with observational constraints. The contours are the $1\sigma$ and $2\sigma$ observational constraints from the  P-ACT-LB data  combined with B-mode measurements from the BICEP and Keck telescopes at the South Pole (BK18),  referred to as P-ACT-LB-BK18. (Right) Corresponding parameter values of $\xi$ and $k$ consistent with observational bounds. The red and blue color region are the constraints from $1\sigma$ and $2\sigma$ observational constraints, respectively.}
    \label{fig:nsr}
\end{figure*}

To examine the general behavior of non-minimal coupling inflation models \eqref{attractor:act}, 
we consider the power-law form,
\begin{equation}\label{power_law:from}
    f(\phi) = \phi^k
\end{equation}
with parameters $\xi > 0$ and $k > 0$, where $k$ is not restricted to integer values. 
For general $k$ and $\xi$, analytic expressions for $n_s$ and $r$ are difficult to obtain, 
so we compute these observables numerically with slow-roll approximation.
In the slow-roll approximation, the scalar spectral index and tensor-to-scalar ratio can be evaluated in the Einstein frame as follows,  
\begin{equation}\label{nsr:sl}
    n_s = 1 - 6 \epsilon_U(\psi) +2 \eta_U(\psi), \quad r =16 \epsilon_U(\psi),
\end{equation}
where the slow-roll parameters are defined by
\begin{equation}
    \epsilon_U(\psi)  = \frac{1}{2} \left(\frac{ U'(\psi)}{U(\psi)}\right)^2, \quad \eta_U(\psi) = \frac{U''(\psi)}{U(\psi)},
\end{equation}
with $U'(\psi) = dU(\psi)/d\psi$ and $U''(\psi) = d^2 U(\psi)/d\psi^2$.
The values of scalar field at the end of inflation $\psi_e$ and at the horizon exit $\psi_*$ are determined by the conditions 
\begin{equation}
    \epsilon_U(\psi_e)=1, \quad N = -\int_{\psi_*}^ {\psi_e} \frac{U(\psi)}{ U'(\psi)}  d\psi.
\end{equation}
Due to the complexity in obtaining the functional relation between  $\psi$ and original $\phi$, the explicit form of the potential $U(\psi)$ in Einstein frame is generally difficult to derive.  To numerically compute the scalar spectral index and the tensor-to-scalar ratio, we recast Eq.  \eqref{nsr:sl} in terms of the scalar field $\phi$ and potential $V_J(\phi)$ in the Jordan frame, using the conformal transformation relations  \eqref{conftransf2} and \eqref{potential:rel}. By adopting the power-law form  \eqref{power_law:from}, the slow-roll parameters become
\begin{equation}\label{sl_eps}
      \epsilon_U[\psi(\phi)]  =\frac{4 k^2}{3 k^2 \xi ^2 \phi ^{2 k} +2 \xi  \phi ^{k+2}+2 \phi ^2}, 
\end{equation}
\begin{equation}
\begin{aligned}\label{sl_eta}
    \eta_U[\psi(\phi)] = & - \frac{4 k \left(3 k^3 \xi ^3 \phi ^{3 k}+(k+2) \xi ^2 \phi ^{2 k+2}+(2-4 k) \phi ^2\right)}{\left(3 k^2 \xi ^2 \phi ^{2 k}+2 \xi  \phi ^{k+2}+2 \phi ^2\right)^2}\\
    & +\frac{4 k \xi  \phi ^k \left(3 k^3 \xi  \phi ^k+(3 k-4) \phi ^2\right)}{\left(3 k^2 \xi ^2 \phi ^{2 k}+2 \xi  \phi ^{k+2}+2 \phi ^2\right)^2}.
\end{aligned}
\end{equation}
The corresponding field value at horizon exit   $\phi_*$ is determined by the number of  $e$-folds:
\begin{equation} \label{efold_jordan}
 N =  \left. \frac{3 k \xi  \phi ^k-3 k \log \left(\xi  \phi ^k+1\right)+\phi ^2}{4 k}\right|_{\phi_e}^{\phi_*},
\end{equation}
where  $\phi_e$ is determined by the condition $\epsilon_U[\psi(\phi_e)]=1$. 
By using  Eqs.  \eqref{sl_eps} - \eqref{efold_jordan}, and setting $N=60$, the scalar spectral index and tensor-to-scalar ratio can be  numerically calculated. The results are shown in Fig.~\ref{fig:nsr}. 
The left panel displays the predictions for $n_s$ and $r$, 
while the right panel shows the corresponding values of $\xi$ and $k$. 
The regions with red and blue points are the constraints from the $1\sigma$ and $2\sigma$ of the observational data, respectively.  
To be consistent with the $1\sigma$ constraint of the observational data, 
the coupling constant $\xi$ and the coupling function index $k$ should satisfy
\begin{equation}
   0.1\lesssim \xi\lesssim35, \quad  0\lesssim k\lesssim 1.5,
\end{equation}
therefore the Higgs inflation, where $k=2$, is ruled out by the $1\sigma$ observational data.  

\subsection{The nonminimal coupling with $\xi \phi^2$}
\begin{figure}
    \includegraphics[width=0.8\linewidth]{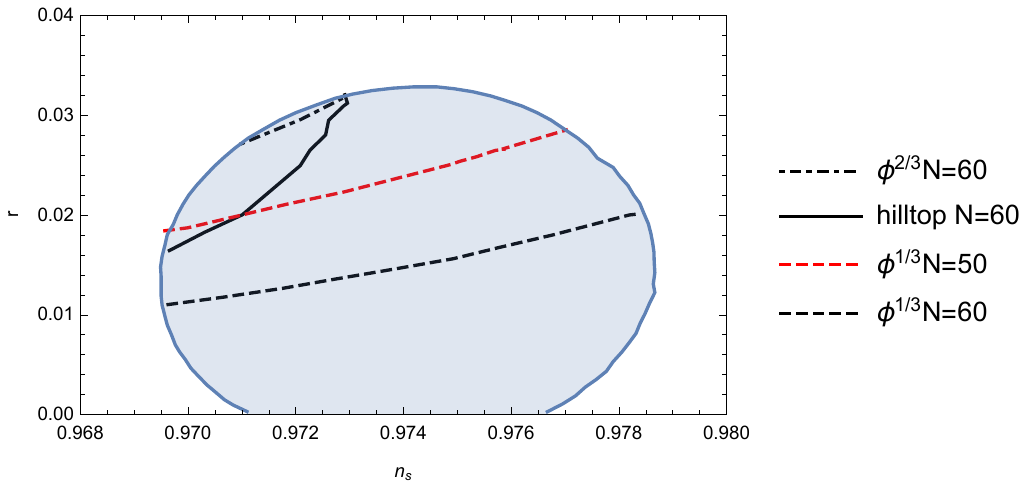}
    \caption{Scalar spectral index $n_s$ and tensor-to-scalar ratio $r$ for different models with the nonminimal coupling $\Omega(\phi)=1+\xi \phi^2$. The contour is the $1\sigma$ observational constraints from the P-ACT-LB-BK18 data. The various colored lines are for different models with varying $\xi$.}
    \label{act2}
\end{figure}

In this subsection, 
we consider the hilltop inflation and power-law inflation with commonly discussed nonminimal coupling $\Omega(\phi)=1+\xi \phi^2$ and small coupling constant $\xi\ll 1$. 
We numerically solve the background and perturbation equations without employing the slow-roll approximation,
and the numerical results of $n_s$ and $r$ are shown in Fig. \ref{act2}.

For the hilltop inflation, the potential is
\begin{equation}
\label{vhilltop}
V(\phi)= V_0\left[1-\left(\frac{\phi}{\mu}\right)^p\right],
\end{equation}
we choose $p=2$ and $\mu=50$, so that $n_s=0.9743$ at $N=60$ for the minimal coupling case where $\xi=0$. 
As shown in Fig. \ref{act2}, for $N=60$,
the constraint on $\xi$ from the $1\sigma$ $n_s-r$ contour, obtained by combining P-ACT-LB data with B-mode measurements from the BICEP and Keck telescopes at the South Pole (BK18) \cite{BICEP:2021xfz}, hereafter referred to as P-ACT-LB-BK18, is $-0.0003<\xi<-0.0002$. 
Unfortunately, for $N=50$, we cannot find any value of $\xi$ to be consistent with the $1\sigma$ $n_s-r$ contour from P-ACT-LB-BK18.

For the power-law inflation, we consider the potential $V(\phi)= V_0\phi^{2/3}$ and $V(\phi)= V_0\phi^{1/3}$.
For the potential $V(\phi)= V_0\phi^{2/3}$, again nonminimal coupling cannot help the model to be consistent with the P-ACT-LB-BK18 constraint at the $1\sigma$ level for $N=50$. 
However, for $N=60$, we find that $0.0005<\xi<0.00073$ at the $1\sigma$ level as shown in Fig. \ref{act2}.

For the pow-law inflation with $V(\phi)= V_0\phi^{1/3}$,
with the help of the nonminimal coupling, the model can satisfy the $1\sigma$ constraints from P-ACT-LB-BK18 as shown in Fig. \ref{act2}.
We find that $-0.00012<\xi<0.00063$ and $0.00014<\xi<0.001$ for $N=50$ and $N=60$, respectively.

\section{conclusion}
The latest observational data from the Atacama Cosmology Telescope have provided new constraints on the scalar spectral index $n_s$. 
The P-ACT-LB results derive a higher value of $n_s$ compared to previous {\it Planck} results.  
This shift in $n_s$ imposes a big challenge on the parameterization of the scalar spectral index, $n_s = 1 - 2/N$, 
which is a universal attractor for a wide range of inflationary models, such as hilltop inflation, T model, E model, 
the Starobinsky model, and inflationary models with a nonminimal coupling $\Omega(\phi)=1+\xi f(\phi)$ and the potential $V_J(\phi)=\lambda f(\phi)^2$  under the strong coupling limit condition $\Omega(\phi)\ll 3\Omega'(\phi)^2/2$. 

However, the attractors with nonminimal coupling are achieved in the strong coupling limit. 
If the coupling constant is small, the attractors are not reached, it is still possible to satisfy the recent P-ACT-LB constraints because a nonminimal coupling helps to reduce the tensor-to-scalar ratio.
We find that with small coupling constant, the predictions for the scalar spectral index $n_s$ and tensor-to-scalar ratio $r$ in inflationary models with nonminimal coupling can still be consistent with the latest P-ACT-LB data at the $1\sigma$ level.  
We consider the general coupling function $\Omega(\phi)=1+\xi f(\phi)$ and the potential $V_J(\phi)=\lambda f^2(\phi)$. 
For the monomial case with $f(\phi)=\phi^k$, to satisfy the  $1\sigma$ region of the observational data, 
the parameter values of $\xi$ and $k$ should satisfy $0.1\lesssim \xi\lesssim35$ and $0\lesssim k\lesssim1.5$.   

We also consider the commonly used nonminimal coupling $\Omega(\phi)=1+\xi \phi^2$.
For the hilltop inflation, the P-ACT-LB-BK18 $1\sigma$ results give the constraint $-0.0003<\xi<-0.0002$ for $N=60$.
For the power-law potential with the power index $2/3$, the P-ACT-LB-BK18 $1\sigma$ results give the constraint $0.0005<\xi<0.00073$ for $N=60$.
For the power-law potential with the power index $1/3$, we find that the P-ACT-LB-BK18 $1\sigma$ contour gives the constraint 
$-0.00012<\xi<0.00063$ and $0.00014<\xi<0.001$ for $N=50$ and $N=60$, respectively. 

\begin{acknowledgments}
This work is supported in part by the National Key Research and Development Program of China under Grant No. 2020YFC2201504, the National Natural Science Foundation of China under Grant No. 12175184, No. 12205015, No. 12305075, the supporting fund for Young Researcher of Beijing Normal University under Grant No. 28719/310432102, the Chongqing Natural Science Foundation under Grant No. CSTB2022NSCQ-MSX1324 and the Startup Research Fund of Henan Academy of Science under Grant No. 241841223.
\end{acknowledgments}


\begin{thebibliography}{25}%
\makeatletter
\providecommand \@ifxundefined [1]{%
 \@ifx{#1\undefined}
}%
\providecommand \@ifnum [1]{%
 \ifnum #1\expandafter \@firstoftwo
 \else \expandafter \@secondoftwo
 \fi
}%
\providecommand \@ifx [1]{%
 \ifx #1\expandafter \@firstoftwo
 \else \expandafter \@secondoftwo
 \fi
}%
\providecommand \natexlab [1]{#1}%
\providecommand \enquote  [1]{``#1''}%
\providecommand \bibnamefont  [1]{#1}%
\providecommand \bibfnamefont [1]{#1}%
\providecommand \citenamefont [1]{#1}%
\providecommand \href@noop [0]{\@secondoftwo}%
\providecommand \href [0]{\begingroup \@sanitize@url \@href}%
\providecommand \@href[1]{\@@startlink{#1}\@@href}%
\providecommand \@@href[1]{\endgroup#1\@@endlink}%
\providecommand \@sanitize@url [0]{\catcode `\\12\catcode `\$12\catcode `\&12\catcode `\#12\catcode `\^12\catcode `\_12\catcode `\%12\relax}%
\providecommand \@@startlink[1]{}%
\providecommand \@@endlink[0]{}%
\providecommand \url  [0]{\begingroup\@sanitize@url \@url }%
\providecommand \@url [1]{\endgroup\@href {#1}{\urlprefix }}%
\providecommand \urlprefix  [0]{URL }%
\providecommand \Eprint [0]{\href }%
\providecommand \doibase [0]{https://doi.org/}%
\providecommand \selectlanguage [0]{\@gobble}%
\providecommand \bibinfo  [0]{\@secondoftwo}%
\providecommand \bibfield  [0]{\@secondoftwo}%
\providecommand \translation [1]{[#1]}%
\providecommand \BibitemOpen [0]{}%
\providecommand \bibitemStop [0]{}%
\providecommand \bibitemNoStop [0]{.\EOS\space}%
\providecommand \EOS [0]{\spacefactor3000\relax}%
\providecommand \BibitemShut  [1]{\csname bibitem#1\endcsname}%
\let\auto@bib@innerbib\@empty
\bibitem [{\citenamefont {Starobinsky}(1980)}]{Starobinsky:1980te}%
  \BibitemOpen
  \bibfield  {author} {\bibinfo {author} {\bibfnamefont {A.~A.}\ \bibnamefont {Starobinsky}},\ }\bibfield  {title} {\bibinfo {title} {{A New Type of Isotropic Cosmological Models Without Singularity}},\ }\href {https://doi.org/10.1016/0370-2693(80)90670-X} {\bibfield  {journal} {\bibinfo  {journal} {Phys. Lett. B}\ }\textbf {\bibinfo {volume} {91}},\ \bibinfo {pages} {99} (\bibinfo {year} {1980})}\BibitemShut {NoStop}%
\bibitem [{\citenamefont {Guth}(1981)}]{Guth:1980zm}%
  \BibitemOpen
  \bibfield  {author} {\bibinfo {author} {\bibfnamefont {A.~H.}\ \bibnamefont {Guth}},\ }\bibfield  {title} {\bibinfo {title} {{The Inflationary Universe: A Possible Solution to the Horizon and Flatness Problems}},\ }\href {https://doi.org/10.1103/PhysRevD.23.347} {\bibfield  {journal} {\bibinfo  {journal} {Phys. Rev. D}\ }\textbf {\bibinfo {volume} {23}},\ \bibinfo {pages} {347} (\bibinfo {year} {1981})}\BibitemShut {NoStop}%
\bibitem [{\citenamefont {Linde}(1982)}]{Linde:1981mu}%
  \BibitemOpen
  \bibfield  {author} {\bibinfo {author} {\bibfnamefont {A.~D.}\ \bibnamefont {Linde}},\ }\bibfield  {title} {\bibinfo {title} {{A New Inflationary Universe Scenario: A Possible Solution of the Horizon, Flatness, Homogeneity, Isotropy and Primordial Monopole Problems}},\ }\href {https://doi.org/10.1016/0370-2693(82)91219-9} {\bibfield  {journal} {\bibinfo  {journal} {Phys. Lett. B}\ }\textbf {\bibinfo {volume} {108}},\ \bibinfo {pages} {389} (\bibinfo {year} {1982})}\BibitemShut {NoStop}%
\bibitem [{\citenamefont {Albrecht}\ and\ \citenamefont {Steinhardt}(1982)}]{Albrecht:1982wi}%
  \BibitemOpen
  \bibfield  {author} {\bibinfo {author} {\bibfnamefont {A.}~\bibnamefont {Albrecht}}\ and\ \bibinfo {author} {\bibfnamefont {P.~J.}\ \bibnamefont {Steinhardt}},\ }\bibfield  {title} {\bibinfo {title} {{Cosmology for Grand Unified Theories with Radiatively Induced Symmetry Breaking}},\ }\href {https://doi.org/10.1103/PhysRevLett.48.1220} {\bibfield  {journal} {\bibinfo  {journal} {Phys. Rev. Lett.}\ }\textbf {\bibinfo {volume} {48}},\ \bibinfo {pages} {1220} (\bibinfo {year} {1982})}\BibitemShut {NoStop}%
\bibitem [{\citenamefont {Boubekeur}\ and\ \citenamefont {Lyth}(2005)}]{Boubekeur:2005zm}%
  \BibitemOpen
  \bibfield  {author} {\bibinfo {author} {\bibfnamefont {L.}~\bibnamefont {Boubekeur}}\ and\ \bibinfo {author} {\bibfnamefont {D.~H.}\ \bibnamefont {Lyth}},\ }\bibfield  {title} {\bibinfo {title} {{Hilltop inflation}},\ }\href {https://doi.org/10.1088/1475-7516/2005/07/010} {\bibfield  {journal} {\bibinfo  {journal} {JCAP}\ }\textbf {\bibinfo {volume} {07}},\ \bibinfo {pages} {010}},\ \Eprint {https://arxiv.org/abs/hep-ph/0502047} {arXiv:hep-ph/0502047} \BibitemShut {NoStop}%
\bibitem [{\citenamefont {Kallosh}\ and\ \citenamefont {Linde}(2013{\natexlab{a}})}]{Kallosh:2013hoa}%
  \BibitemOpen
  \bibfield  {author} {\bibinfo {author} {\bibfnamefont {R.}~\bibnamefont {Kallosh}}\ and\ \bibinfo {author} {\bibfnamefont {A.}~\bibnamefont {Linde}},\ }\bibfield  {title} {\bibinfo {title} {{Universality Class in Conformal Inflation}},\ }\href {https://doi.org/10.1088/1475-7516/2013/07/002} {\bibfield  {journal} {\bibinfo  {journal} {JCAP}\ }\textbf {\bibinfo {volume} {07}},\ \bibinfo {pages} {002}},\ \Eprint {https://arxiv.org/abs/1306.5220} {arXiv:1306.5220 [hep-th]} \BibitemShut {NoStop}%
\bibitem [{\citenamefont {Kallosh}\ and\ \citenamefont {Linde}(2013{\natexlab{b}})}]{Kallosh:2013maa}%
  \BibitemOpen
  \bibfield  {author} {\bibinfo {author} {\bibfnamefont {R.}~\bibnamefont {Kallosh}}\ and\ \bibinfo {author} {\bibfnamefont {A.}~\bibnamefont {Linde}},\ }\bibfield  {title} {\bibinfo {title} {{Non-minimal Inflationary Attractors}},\ }\href {https://doi.org/10.1088/1475-7516/2013/10/033} {\bibfield  {journal} {\bibinfo  {journal} {JCAP}\ }\textbf {\bibinfo {volume} {10}},\ \bibinfo {pages} {033}},\ \Eprint {https://arxiv.org/abs/1307.7938} {arXiv:1307.7938 [hep-th]} \BibitemShut {NoStop}%
\bibitem [{\citenamefont {Kaiser}(1995)}]{Kaiser:1994vs}%
  \BibitemOpen
  \bibfield  {author} {\bibinfo {author} {\bibfnamefont {D.~I.}\ \bibnamefont {Kaiser}},\ }\bibfield  {title} {\bibinfo {title} {{Primordial spectral indices from generalized Einstein theories}},\ }\href {https://doi.org/10.1103/PhysRevD.52.4295} {\bibfield  {journal} {\bibinfo  {journal} {Phys. Rev. D}\ }\textbf {\bibinfo {volume} {52}},\ \bibinfo {pages} {4295} (\bibinfo {year} {1995})},\ \Eprint {https://arxiv.org/abs/astro-ph/9408044} {arXiv:astro-ph/9408044} \BibitemShut {NoStop}%
\bibitem [{\citenamefont {Bezrukov}\ and\ \citenamefont {Shaposhnikov}(2008)}]{Bezrukov:2007ep}%
  \BibitemOpen
  \bibfield  {author} {\bibinfo {author} {\bibfnamefont {F.~L.}\ \bibnamefont {Bezrukov}}\ and\ \bibinfo {author} {\bibfnamefont {M.}~\bibnamefont {Shaposhnikov}},\ }\bibfield  {title} {\bibinfo {title} {{The Standard Model Higgs boson as the inflaton}},\ }\href {https://doi.org/10.1016/j.physletb.2007.11.072} {\bibfield  {journal} {\bibinfo  {journal} {Phys. Lett. B}\ }\textbf {\bibinfo {volume} {659}},\ \bibinfo {pages} {703} (\bibinfo {year} {2008})},\ \Eprint {https://arxiv.org/abs/0710.3755} {arXiv:0710.3755 [hep-th]} \BibitemShut {NoStop}%
\bibitem [{\citenamefont {Akrami}\ \emph {et~al.}(2020)\citenamefont {Akrami} \emph {et~al.}}]{Planck:2018jri}%
  \BibitemOpen
  \bibfield  {author} {\bibinfo {author} {\bibfnamefont {Y.}~\bibnamefont {Akrami}} \emph {et~al.} (\bibinfo {collaboration} {Planck}),\ }\bibfield  {title} {\bibinfo {title} {{Planck 2018 results. X. Constraints on inflation}},\ }\href {https://doi.org/10.1051/0004-6361/201833887} {\bibfield  {journal} {\bibinfo  {journal} {Astron. Astrophys.}\ }\textbf {\bibinfo {volume} {641}},\ \bibinfo {pages} {A10} (\bibinfo {year} {2020})},\ \Eprint {https://arxiv.org/abs/1807.06211} {arXiv:1807.06211 [astro-ph.CO]} \BibitemShut {NoStop}%
\bibitem [{\citenamefont {Kallosh}\ \emph {et~al.}(2014)\citenamefont {Kallosh}, \citenamefont {Linde},\ and\ \citenamefont {Roest}}]{Kallosh:2013tua}%
  \BibitemOpen
  \bibfield  {author} {\bibinfo {author} {\bibfnamefont {R.}~\bibnamefont {Kallosh}}, \bibinfo {author} {\bibfnamefont {A.}~\bibnamefont {Linde}},\ and\ \bibinfo {author} {\bibfnamefont {D.}~\bibnamefont {Roest}},\ }\bibfield  {title} {\bibinfo {title} {{Universal Attractor for Inflation at Strong Coupling}},\ }\href {https://doi.org/10.1103/PhysRevLett.112.011303} {\bibfield  {journal} {\bibinfo  {journal} {Phys. Rev. Lett.}\ }\textbf {\bibinfo {volume} {112}},\ \bibinfo {pages} {011303} (\bibinfo {year} {2014})},\ \Eprint {https://arxiv.org/abs/1310.3950} {arXiv:1310.3950 [hep-th]} \BibitemShut {NoStop}%
\bibitem [{\citenamefont {Louis}\ \emph {et~al.}(2025)\citenamefont {Louis} \emph {et~al.}}]{Louis:2025tst}%
  \BibitemOpen
  \bibfield  {author} {\bibinfo {author} {\bibfnamefont {T.}~\bibnamefont {Louis}} \emph {et~al.} (\bibinfo {collaboration} {ACT}),\ }\bibfield  {title} {\bibinfo {title} {{The Atacama Cosmology Telescope: DR6 Power Spectra, Likelihoods and $\Lambda$CDM Parameters}},\ }\href@noop {} {\  (\bibinfo {year} {2025})},\ \Eprint {https://arxiv.org/abs/2503.14452} {arXiv:2503.14452 [astro-ph.CO]} \BibitemShut {NoStop}%
\bibitem [{\citenamefont {Calabrese}\ \emph {et~al.}(2025)\citenamefont {Calabrese} \emph {et~al.}}]{ACT:2025tim}%
  \BibitemOpen
  \bibfield  {author} {\bibinfo {author} {\bibfnamefont {E.}~\bibnamefont {Calabrese}} \emph {et~al.} (\bibinfo {collaboration} {ACT}),\ }\bibfield  {title} {\bibinfo {title} {{The Atacama Cosmology Telescope: DR6 Constraints on Extended Cosmological Models}},\ }\href@noop {} {\  (\bibinfo {year} {2025})},\ \Eprint {https://arxiv.org/abs/2503.14454} {arXiv:2503.14454 [astro-ph.CO]} \BibitemShut {NoStop}%
\bibitem [{\citenamefont {Adame}\ \emph {et~al.}(2025{\natexlab{a}})\citenamefont {Adame} \emph {et~al.}}]{DESI:2024uvr}%
  \BibitemOpen
  \bibfield  {author} {\bibinfo {author} {\bibfnamefont {A.~G.}\ \bibnamefont {Adame}} \emph {et~al.} (\bibinfo {collaboration} {DESI}),\ }\bibfield  {title} {\bibinfo {title} {{DESI 2024 III: baryon acoustic oscillations from galaxies and quasars}},\ }\href {https://doi.org/10.1088/1475-7516/2025/04/012} {\bibfield  {journal} {\bibinfo  {journal} {JCAP}\ }\textbf {\bibinfo {volume} {04}},\ \bibinfo {pages} {012}},\ \Eprint {https://arxiv.org/abs/2404.03000} {arXiv:2404.03000 [astro-ph.CO]} \BibitemShut {NoStop}%
\bibitem [{\citenamefont {Adame}\ \emph {et~al.}(2025{\natexlab{b}})\citenamefont {Adame} \emph {et~al.}}]{DESI:2024mwx}%
  \BibitemOpen
  \bibfield  {author} {\bibinfo {author} {\bibfnamefont {A.~G.}\ \bibnamefont {Adame}} \emph {et~al.} (\bibinfo {collaboration} {DESI}),\ }\bibfield  {title} {\bibinfo {title} {{DESI 2024 VI: cosmological constraints from the measurements of baryon acoustic oscillations}},\ }\href {https://doi.org/10.1088/1475-7516/2025/02/021} {\bibfield  {journal} {\bibinfo  {journal} {JCAP}\ }\textbf {\bibinfo {volume} {02}},\ \bibinfo {pages} {021}},\ \Eprint {https://arxiv.org/abs/2404.03002} {arXiv:2404.03002 [astro-ph.CO]} \BibitemShut {NoStop}%
\bibitem [{\citenamefont {Kallosh}\ \emph {et~al.}(2025)\citenamefont {Kallosh}, \citenamefont {Linde},\ and\ \citenamefont {Roest}}]{Kallosh:2025rni}%
  \BibitemOpen
  \bibfield  {author} {\bibinfo {author} {\bibfnamefont {R.}~\bibnamefont {Kallosh}}, \bibinfo {author} {\bibfnamefont {A.}~\bibnamefont {Linde}},\ and\ \bibinfo {author} {\bibfnamefont {D.}~\bibnamefont {Roest}},\ }\bibfield  {title} {\bibinfo {title} {{A simple scenario for the last ACT}},\ }\href@noop {} {\  (\bibinfo {year} {2025})},\ \Eprint {https://arxiv.org/abs/2503.21030} {arXiv:2503.21030 [hep-th]} \BibitemShut {NoStop}%
\bibitem [{\citenamefont {Gialamas}\ \emph {et~al.}(2025)\citenamefont {Gialamas}, \citenamefont {Karam}, \citenamefont {Racioppi},\ and\ \citenamefont {Raidal}}]{Gialamas:2025kef}%
  \BibitemOpen
  \bibfield  {author} {\bibinfo {author} {\bibfnamefont {I.~D.}\ \bibnamefont {Gialamas}}, \bibinfo {author} {\bibfnamefont {A.}~\bibnamefont {Karam}}, \bibinfo {author} {\bibfnamefont {A.}~\bibnamefont {Racioppi}},\ and\ \bibinfo {author} {\bibfnamefont {M.}~\bibnamefont {Raidal}},\ }\bibfield  {title} {\bibinfo {title} {{Has ACT measured radiative corrections to the tree-level Higgs-like inflation?}},\ }\href@noop {} {\  (\bibinfo {year} {2025})},\ \Eprint {https://arxiv.org/abs/2504.06002} {arXiv:2504.06002 [astro-ph.CO]} \BibitemShut {NoStop}%
\bibitem [{\citenamefont {Fr\"ob}\ \emph {et~al.}(2025)\citenamefont {Fr\"ob}, \citenamefont {Glavan}, \citenamefont {Meda},\ and\ \citenamefont {Sawicki}}]{Frob:2025sfq}%
  \BibitemOpen
  \bibfield  {author} {\bibinfo {author} {\bibfnamefont {M.~B.}\ \bibnamefont {Fr\"ob}}, \bibinfo {author} {\bibfnamefont {D.}~\bibnamefont {Glavan}}, \bibinfo {author} {\bibfnamefont {P.}~\bibnamefont {Meda}},\ and\ \bibinfo {author} {\bibfnamefont {I.}~\bibnamefont {Sawicki}},\ }\bibfield  {title} {\bibinfo {title} {{One-loop correction to primordial tensor modes during radiation era}},\ }\href@noop {} {\  (\bibinfo {year} {2025})},\ \Eprint {https://arxiv.org/abs/2504.02609} {arXiv:2504.02609 [astro-ph.CO]} \BibitemShut {NoStop}%
\bibitem [{\citenamefont {Dioguardi}\ \emph {et~al.}(2025)\citenamefont {Dioguardi}, \citenamefont {Iovino},\ and\ \citenamefont {Racioppi}}]{Dioguardi:2025vci}%
  \BibitemOpen
  \bibfield  {author} {\bibinfo {author} {\bibfnamefont {C.}~\bibnamefont {Dioguardi}}, \bibinfo {author} {\bibfnamefont {A.~J.}\ \bibnamefont {Iovino}},\ and\ \bibinfo {author} {\bibfnamefont {A.}~\bibnamefont {Racioppi}},\ }\bibfield  {title} {\bibinfo {title} {{Fractional attractors in light of the latest ACT observations}},\ }\href@noop {} {\  (\bibinfo {year} {2025})},\ \Eprint {https://arxiv.org/abs/2504.02809} {arXiv:2504.02809 [gr-qc]} \BibitemShut {NoStop}%
\bibitem [{\citenamefont {Brahma}\ and\ \citenamefont {Calder\'on-Figueroa}(2025)}]{Brahma:2025dio}%
  \BibitemOpen
  \bibfield  {author} {\bibinfo {author} {\bibfnamefont {S.}~\bibnamefont {Brahma}}\ and\ \bibinfo {author} {\bibfnamefont {J.}~\bibnamefont {Calder\'on-Figueroa}},\ }\bibfield  {title} {\bibinfo {title} {{Is the CMB revealing signs of pre-inflationary physics?}},\ }\href@noop {} {\  (\bibinfo {year} {2025})},\ \Eprint {https://arxiv.org/abs/2504.02746} {arXiv:2504.02746 [astro-ph.CO]} \BibitemShut {NoStop}%
\bibitem [{\citenamefont {Berera}\ \emph {et~al.}(2025)\citenamefont {Berera}, \citenamefont {Brahma}, \citenamefont {Qiu}, \citenamefont {O.~Ramos},\ and\ \citenamefont {Rodrigues}}]{Berera:2025vsu}%
  \BibitemOpen
  \bibfield  {author} {\bibinfo {author} {\bibfnamefont {A.}~\bibnamefont {Berera}}, \bibinfo {author} {\bibfnamefont {S.}~\bibnamefont {Brahma}}, \bibinfo {author} {\bibfnamefont {Z.}~\bibnamefont {Qiu}}, \bibinfo {author} {\bibfnamefont {R.}~\bibnamefont {O.~Ramos}},\ and\ \bibinfo {author} {\bibfnamefont {G.~S.}\ \bibnamefont {Rodrigues}},\ }\bibfield  {title} {\bibinfo {title} {{The early universe is $\textit{ACT}$-ing $\textit{warm}$}},\ }\href@noop {} {\  (\bibinfo {year} {2025})},\ \Eprint {https://arxiv.org/abs/2504.02655} {arXiv:2504.02655 [hep-th]} \BibitemShut {NoStop}%
\bibitem [{\citenamefont {Aoki}\ \emph {et~al.}(2025)\citenamefont {Aoki}, \citenamefont {Otsuka},\ and\ \citenamefont {Yanagita}}]{Aoki:2025wld}%
  \BibitemOpen
  \bibfield  {author} {\bibinfo {author} {\bibfnamefont {S.}~\bibnamefont {Aoki}}, \bibinfo {author} {\bibfnamefont {H.}~\bibnamefont {Otsuka}},\ and\ \bibinfo {author} {\bibfnamefont {R.}~\bibnamefont {Yanagita}},\ }\bibfield  {title} {\bibinfo {title} {{Higgs-Modular Inflation}},\ }\href@noop {} {\  (\bibinfo {year} {2025})},\ \Eprint {https://arxiv.org/abs/2504.01622} {arXiv:2504.01622 [hep-ph]} \BibitemShut {NoStop}%
\bibitem [{\citenamefont {Dioguardi}\ and\ \citenamefont {Karam}(2025)}]{Dioguardi:2025mpp}%
  \BibitemOpen
  \bibfield  {author} {\bibinfo {author} {\bibfnamefont {C.}~\bibnamefont {Dioguardi}}\ and\ \bibinfo {author} {\bibfnamefont {A.}~\bibnamefont {Karam}},\ }\bibfield  {title} {\bibinfo {title} {{Palatini Linear Attractors Are Back in ACTion}},\ }\href@noop {} {\  (\bibinfo {year} {2025})},\ \Eprint {https://arxiv.org/abs/2504.12937} {arXiv:2504.12937 [gr-qc]} \BibitemShut {NoStop}%
\bibitem [{\citenamefont {Salvio}(2025)}]{Salvio:2025izr}%
  \BibitemOpen
  \bibfield  {author} {\bibinfo {author} {\bibfnamefont {A.}~\bibnamefont {Salvio}},\ }\bibfield  {title} {\bibinfo {title} {{Independent connection in ACTion during inflation}},\ }\href@noop {} {\  (\bibinfo {year} {2025})},\ \Eprint {https://arxiv.org/abs/2504.10488} {arXiv:2504.10488 [hep-ph]} \BibitemShut {NoStop}%
\bibitem [{\citenamefont {Ade}\ \emph {et~al.}(2021)\citenamefont {Ade} \emph {et~al.}}]{BICEP:2021xfz}%
  \BibitemOpen
  \bibfield  {author} {\bibinfo {author} {\bibfnamefont {P.~A.~R.}\ \bibnamefont {Ade}} \emph {et~al.} (\bibinfo {collaboration} {BICEP, Keck}),\ }\bibfield  {title} {\bibinfo {title} {{Improved Constraints on Primordial Gravitational Waves using Planck, WMAP, and BICEP/Keck Observations through the 2018 Observing Season}},\ }\href {https://doi.org/10.1103/PhysRevLett.127.151301} {\bibfield  {journal} {\bibinfo  {journal} {Phys. Rev. Lett.}\ }\textbf {\bibinfo {volume} {127}},\ \bibinfo {pages} {151301} (\bibinfo {year} {2021})},\ \Eprint {https://arxiv.org/abs/2110.00483} {arXiv:2110.00483 [astro-ph.CO]} \BibitemShut {NoStop}%
\end{thebibliography}
%

\end{document}